\documentclass[twocolumn]{emulateapj}
\usepackage{graphicx}
\usepackage{amssymb}
\usepackage{amsmath}
\usepackage{subfigure}
\usepackage{multirow}
\usepackage{threeparttable}
\usepackage{array}
\usepackage{natbib}
\usepackage{xcolor}
\usepackage[colorlinks]{hyperref}

\def\MgX{{\rm Mg} {\sc x}}
\def\NaIX{{\rm Na} {\sc ix}}
\def\NeVIII{{\rm Ne} {\sc viii}}
\def\NeV{{\rm Ne} {\sc v}}
\def\NeVI{{\rm Ne} {\sc vi}}
\def\NeIV{{\rm Ne} {\sc iv}}
\def\OIII{{\rm O} {\sc iii}}
\def\OIV{{\rm O} {\sc iv}}
\def\OV{{\rm O} {\sc v}}
\def\OVI{{\rm O} {\sc vi}}
\def\OVII{{\rm O} {\sc vii}}
\def\OVIII{{\rm O} {\sc viii}}
\def\NIV{{\rm N} {\sc iv}}
\def\FeII{{\rm Fe} {\sc ii}}
\def\SiII{{\rm Si} {\sc ii}}
\def\SiIV{{\rm Si} {\sc iv}}
\def\CIII{{\rm C} {\sc iii}}
\def\HI{{\rm H} {\sc i}}
\def\NiII{{\rm Ni} {\sc ii}}

\def\kms{~\rm km~s^{-1}}      
\def\cmsq{~\rm cm^{-2}}
\def\cc{~\rm cm^{-3}}

\begin{document}
\title{A hot gaseous galaxy halo candidate with {\MgX} absorption}
\author{Zhijie Qu and Joel N. Bregman}
\affil{Department of Astronomy, University of Michigan, Ann Arbor, MI 48104, USA}
\email{quzhijie@umich.edu}
\email{jbregman@umich.edu}

\begin{abstract}
The hot gas in galaxy halos may account for a significant fraction of missing baryons in galaxies, and some of these gases can be traced by high ionization absorption systems in QSO UV spectra. Using high S/N ratio {\it HST}/COS spectra, we discovered a high ionization state system at $z=1.1912$ in the sightline toward \object{LBQS 1435-0134}, and two-components absorption lines are matched for {\MgX}, {\NeVIII}, {\NeVI}, {\OVI}, {\NeV}, {\OV}, {\NeIV}, {\OIV}, {\NIV}, {\OIII}, and {\HI}. {\MgX}, detected for the first time ($5.8 \sigma$), is a particularly direct tracer of hot galactic halos, as its peak ion fraction occurs near $10^{6.1}\rm~ K$, about the temperature of a virialized hot galaxy halo of mass $\sim 0.5 M^*$. With {\MgX} and {\NeVIII}, a photoionization model cannot reproduce the observed column densities with path lengths of galaxy halos. For collisional ionization models, one or two temperature models do not produce acceptable fits, but a three temperature model or a power law model can produce the observed results. In the power law model, ${\rm d}N/{\rm d}T = 10^{4.4\pm 2.2-[Z/X]} T^{1.55\pm 0.41}$ with temperatures in the range $10^{4.39\pm0.13} {\rm~K} < T < 10^{6.04\pm0.05}~ {\rm K}$, the total hydrogen column density is $8.2 \times 10^{19} (0.3Z_{\odot}/Z) \cmsq$ and the positive power law index indicates most of the mass is at the high temperature end. We suggest that this absorption system is a hot volume-filled galaxy halo rather than interaction layers between the hot halo and cool clouds. The temperature dependence of the column density is likely due to the local mixture of multiple phase gases.
\end{abstract} 
\keywords{galaxies: halos -- galaxies: ISM -- quasars: absorption lines}
\maketitle

\section{Introduction}
\label{intro}
The baryon masses of galaxies are significantly smaller than expectations based on the cosmological baryon to total mass ratio ($0.16$; \citealt{Planck-Collaboration:2015aa}) where the total mass is inferred from the galaxy rotation curve. In particular, the stellar mass and the galactic gas mass can only account for $\sim 15-25\%$ of the total baryon mass \citep{Mateo:1998aa, Bell:2003aa, Dai:2010aa, Martin:2010aa, Anderson:2010aa, Behroozi:2010aa, McGaugh:2015aa}. A solution to this missing baryon problem in galaxies is that there exists a massive extended hot gaseous halo \citep{Fukugita:2006aa, Bregman:2007ab, Kaufmann:2009aa}.

The gaseous content of halos is predicted to evolve during galaxy formation and evolution. Early in the history of the universe ($z\sim 4$), baryons are expected to be cool ($\sim 10^4\rm~ K$; \citealt{Weinberg:1997aa}), while during galaxy formation ($z\sim 3$ to $1$), the cool gas is transformed into the warm-hot medium ($10^5 - 10^7\rm~K$) and the hot medium ($>10^7\rm~K$) in halos as a consequence of the gravitational collapse and the shock heating \citep{Cen:1999aa}. For low redshift galaxies ($z \lesssim 1$), the predicted cool gas accounts for $\sim 20\% - 40\%$ of the total baryons, while more than $30\%$ should be in the warm-hot intergalactic medium (WHIM; \citealt{Cen:2006aa}). The gaseous halo depends on the mass of the galaxy, with extensive hot halos present around massive galaxies but not around low-mass galaxies \citep{Keres:2005aa, Keres:2009aa}.

These theoretical works stimulated observational searches for the halo gas, determining masses and temperatures. The cool halo gas ($\sim 10^3-10^5\rm~K$) can be traced by low ionization metal ion absorption, which is detected in spectra of background active galactic nuclei (AGN; e.g. \citealt{Werk:2014aa}). The COS-Halo team carried out a survey of ultraviolet (UV) absorption lines of low redshift galaxies ($z \approx 0.2$) mainly using the Cosmic Origins Spectrograph (COS; \citealt{Green:2012aa}) on the {\it Hubble Space Telescope} ({\it HST}). They found that $\approx 90\%$ of their target galaxies show cool gas absorption due to galaxy halos \citep{Werk:2012aa, Werk:2013aa, Tumlinson:2013aa}. Based on their observations, the COS-Halo team found the cool gas in the halo can account for $\sim 40 \%$ of the total baryons for $L\approx L^{*}$ galaxies \citep{Werk:2014aa}, while a recent study showed that this component can only account for $\sim 6\%$ \citep{Stern:2016aa}. These cool absorption systems are believed to be separate clouds similar to high velocity clouds in the Milky Way \citep{Sembach:2003aa, Zahedy:2016aa}. In addition to the atomic gas, cooler molecular gas can be present in higher column density systems, such as the damped Ly$\alpha$ absorbers, which could also be hosted by galaxy halos \citep{Srianand:2005aa, Muzahid:2015aa, Muzahid:2016aa}.

Besides the cool gas, the theory predicts the presence of a hot volume-filling medium at about the virial temperature in a galaxy halo, and such gas has been detected. For an $L^{*}$ galaxy, the virial temperature is about $10^{6.3}$ K, so the most prominent emission and absorption lines are in the X-ray band (see \citealt{Bregman:2007aa} for review). X-ray emission line studies of external galaxies detect extended hot halos to $0.1R_{200}$, beyond which the surface brightness falls to undetectable levels, when the gas density is around $n_{200}\sim 5\times10^{-5}\rm~cm^{-3}$ (\citealt{Anderson:2010aa, Boroson:2011aa, Bogdan:2015aa, Anderson:2016aa}). Within $0.1R_{200}$, smaller than $5\%$ of total galactic baryons are detected, and an extrapolation to $R_{200}$ increases this mass by about an order of magnitude, still not accounting for all the baryons. Absorption line studies of intervening gas (hosted by galaxy halos) have not been fruitful due to the short redshift search space \citep{Bregman:2015aa} and to the contamination by weak Galactic lines \citep{Nicastro:2016aa,Nicastro:2016ab}. X-ray studies of the Milky Way utilize both emission lines and absorption lines of {\OVII} and {\OVIII}, and also detect a hot halo extending to at least $50\rm~ kpc$ \citep{Miller:2015aa, Miller:2016aa, Hodges-Kluck:2016aa}. When extrapolated to $R_{200}$, the gaseous mass is $\sim 4.3\times 10^{10}M_{\odot}$ within $250\rm~kpc$, which is comparable to the total stellar mass \citep{Miller:2015aa}.

Another approach focuses on high ionization species in the UV band to detect the hot halo. Typically, low ionization species are ionized by photons from background AGNs. Meanwhile, ions with ionizational potentials greater than {\OVI} ($113.9\rm~eV$) are collisionally ionized in a halo environment, and can be employed to trace the warm-hot gas \citep{Tripp:2008aa, Tepper-Garcia:2013aa, Stocke:2014aa}. The two most useful ions are {\NeVIII} and {\MgX}, which have resonance double lines in the extreme UV band \citep{Verner:1994aa}. The first intergalactic {\NeVIII} absorption feature was discovered by \citet{Savage:2005aa} with the Far Ultraviolet Spectroscopic Explorer (FUSE). Subsequently, several {\NeVIII} doublets have been detected with matched {\OVI} doublets \citep{Tripp:2011aa, Narayanan:2011aa, Meiring:2013aa}. Thus, assuming all {\OVI} ions are also collisional ionized, intergalactic {\NeVIII} systems can be described as a single temperature collisional equilibrium model with temperatures of $\sim 10^{5.7}\rm~K$ and total column densities of $\sim 10^{18.8} - 10^{20.1} \rm~cm^{-2}$ \citep{Narayanan:2012aa}. Although the column density is within the range expected for a volume-filling galaxy halo, this temperature is lower than the expectation ($\gtrsim 10^6\rm~K$), so it is suggested that the detected gas originates from interaction layers, which are the interfaces between cool clouds and the hot halo \citep{Gnat:2010aa, Kwak:2015aa}, rather than the isolated hot halo \citep{Narayanan:2011aa, Meiring:2013aa}. Even for the Milky Way, a multi-temperature model might be more physically relevant, with one component at the virial temperature (as is found for the Milky Way; \citealt{Miller:2015aa}), yet obtaining a unique model fit would require additional data.

One way to solve this dilemma is to expand the probed temperature range by including {\MgX} as a diagnostic, because it reaches a peak ionization at about twice the temperature of {\NeVIII} ($\sim 10^{6.1}\rm~K$, compared to $\sim 10^{5.8}\rm~K$) and four times that of {\OVI} ($\sim 10^{5.5}\rm~K$; \citealt{Bryans:2006aa}). The gas traced by {\MgX} is approximately at the virial temperature for a $\sim L^{*}$ galaxy, making this the ideal diagnostic for studies of hot gaseous halos. A blind search of {\MgX} was carried out with {\it HST}/COS ({\it HST} proposal ID 11741; Tripp as PI). However, currently discovered {\MgX} doublets in their spectra have been identified only from AGN outflows \citep{Muzahid:2013aa}. AGN outflows show several differences from a hot galactic halo, so distinguishing between the two possibilities is viable in some but not all cases. First, AGN outflows show larger observed column densities of {\MgX} than predicted for galactic halos ($\gtrsim 10^{14.5}\rm~cm^{-2}$), and even less abundant species, such as {\NaIX} may be detectable. If one assumed these lines come from gaseous halos, the expected hydrogen column density is $\gtrsim 10^{21}\rm~cm^{-2}$, which is about one order of magnitude higher than the typical column density for the halo of an isolated galaxy ($n_{200}R_{200}\lesssim 10^{20}\rm~cm^{-2}$). Also, density sensitive lines from excited ions (e.g. {\OIV}$^*$) seem to be common in AGN outflows \citep{Finn:2014aa}. These lines trace the gas with a density $\gtrsim 10^{2}\rm~cm^{-3}$, at least $10^5$ greater than galaxy halo densities.

The {\MgX} absorption line equivalent width from a typical galaxy halo would be $20\rm~ m\AA$ in the rest frame ($40~\rm m\AA$ at $z = 1$), which is detectable in a continuum with a S/N $> 20$. In this paper, we report on such a detection in the sightline toward \object{LBQS 1435-0134}, which contains {\MgX}, {\NeVIII}, {\OVI} and a number of other highly ionized ions. We introduce the object and our data reduction methods in Section \ref{method}. Section \ref{LBQS1435} presents the results of absorption systems along this sightline, and the modeling of the {\MgX} system is in Section \ref{model}. We discuss the origin and the implication of this absorption system in Section \ref{discussion}.

\section{Objects and methodology}
\label{method}
We carried out a blind search for {\MgX} doublets using archived COS far-ultraviolet (FUV) spectra in the Mikulski Archive for Space Telescopes (MAST). The {\MgX} doublet occurs at $609.8$ and $624.9\rm~\AA$, so it is only visible when the redshift is larger than $0.92$ in the common COS G130M grating settings with shortest wavelength of $1170\rm~\AA$. The highest search redshift is limited by the QSO redshift. Here, we avoid a $5000\rm~ km~s^{-1}$ velocity region around the QSO redshift to account for intrinsic absorption. Thus, we set the redshift threshold for our selected sample to $z>1$ with the minimum search upper limit of $0.97$ to ensure a search space of $\Delta z \geq 0.05$ for each source.

In this search, we employed the Bayesian Blocks method to characterize the spectra to help us to find absorption features, and several {\MgX} doublets have been discovered. However, most of them are too strong to be associated with galaxy halos, and also have matched density sensitive lines, which should originate from AGN winds. After excluding these AGN outflow absorption systems, we found one absorption system in \object{LBQS 1435-0134} with {\MgX} and several absorption features from Ne and O. Also, \object{LBQS 1435-0134} is the only object with a sufficiently high S/N ratio $> 20$, while other objects are $\lesssim 10$.

\begin{figure*}
\begin{center}
\includegraphics[width=0.95\textwidth]{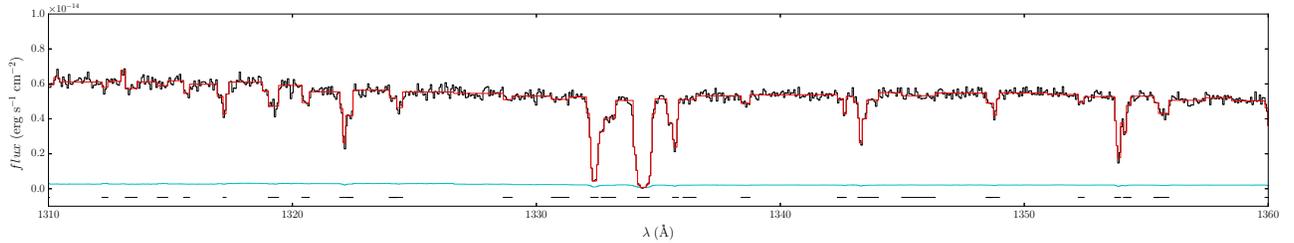}
\end{center}
\caption{This shows an example of the Bayesian Blocks algorithm and the line determination. The black and cyan lines are the rebinned fluxes and their errors from {\it coadd\_x1d.pro}, while the red line is the result of Bayesian Blocks. Adopted absorption lines are shown as black boxes along the bottom.}
\label{f1}
\end{figure*}

\subsection{LBQS 1435-0134}
LBQS 1435-0134 was discovered in the optical survey the Large Bright Quasar Survey (LBQS; \citealt{Hewett:1995aa}). Subsequently, the NRAO VLA Sky Survey (NVSS) showed that it is a radio loud QSO with a flux density of $52.7\pm1.7\rm ~mJy$ at $1.4\rm~GHz$ \citep{Condon:1998aa}, and with a redshift of $1.310790$, as measured in the Sloan Digital Sky Survey (SDSS; \citealt{Hewett:2010aa}). The high S/N ratio UV spectra were obtained by COS and the Space Telescope Imaging Spectrograph (STIS) on {\it HST} in the FUV and the near-ultraviolet (NUV) bands, and the total exposures are $56\rm~ksec$ and $21\rm~ ksec$ for COS/FUV and STIS/MAMA, respectively.

The COS observations were acquired on 2010 August 8 to 22 as a part of {\it HST} program 11741 (PI: Tripp). To achieve the full wavelength coverage of COS/FUV ($1150-1800\rm~ \AA$), two different central wavelengths were used for the two medium resolution gratings (G130M and G160M). For G130M, the central wavelengths $1309\rm~\AA$ and $1327\rm~\AA$ were employed to fill the gap between two detector segments, while for G160M, they were $1600\rm~\AA$ and $1623\rm~\AA$. The total exposures of G130M and G160M gratings were $22\rm~ksec$ and $34\rm~ksec$, respectively, and the times were divided into even parts for two central wavelengths. This configuration of split positions reduces the impact of the fixed pattern noise, which makes the weak line more reliable \citep{Savage:2005aa}. The STIS spectra were obtained on 2015 June 1 to 2 in {\it HST} program 13846 (PI: Tripp). Only one grating (E230M) with a slit of $0.2'\times0.2'$ and one central wavelength ($2415\rm~\AA$) were employed, and the total exposure time was $21\rm~ksec$.

We mainly focused on COS/FUV spectra because of the higher S/N ratio, while STIS spectra were only employed to match the {\OVI} doublet and the Lyman series. Thus, the following search method was only applied to COS data.

\subsection{The COS Spectral Reduction}
The archived {\it x1d} files from MAST are coadded using routines from the COS team ({\it coadd\_x1d.pro}), introduced in \citet{Danforth:2010aa}. Exposures from G130M and G160M are coadded together and resampled in 6 raw pixels ($\approx 0.06 - 0.07$ \AA), which produces a spectral resolution of $\approx 20000$. During this coadd process, strong galactic lines (e.g. {\SiII} $\lambda\lambda~1260.4$) are employed to align exposures. However, this alignment cannot eliminate possible instrumental shifts \citep{Tumlinson:2013aa}. This instrumental shift is caused by the geometric distortion and wavelength solutions of COS, which could be up to $20~\rm km~s^{-1}$ in coadded spectra. For the comparison between two wavelength positions, the instrumental shift could be around $30\rm~km~s^{-1}$, which must be considered in the following profile fitting.

The coadded spectra are characterized using the Bayesian Blocks algorithm to prepare for the absorption feature search and the line identification. Bayesian Blocks was developed to study the time variability in high energy astrophysics, such as gamma-ray bursts, but is applicable to any one-dimensional data set, even if it is discontinuous \citep{Scargle:1998aa, Scargle:2013aa}. To use this algorithm in the spectrum cases, we employed the equivalent count, defined as the square of the S/N ratio for each pixel, for each resolution element as the input of Bayesian Blocks. As a non-parametric method, this algorithm represents the spectrum as a series of wavelength intervals, generally unequal in length (blocks), within which the flux is modeled as a constant. These blocks are extracted based on the change points (the beginning or the end points), which is determined by well-tested statistical measures, thus the end result is a spectrum represented by a set of step functions. A reduction example is given in Fig. \ref{f1}.

With the output from the Bayesian Blocks algorithm, absorption lines can be extracted by analyzing the structure of ``Bayesian blocks". Absorption blocks are defined as blocks that are local nadirs and shorter than 30 bins in coadded spectra ($\approx 1.8\rm~\AA$ for G130M and $\approx 2.2\rm~\AA$ for G160M). This width ($\approx 2\rm~\AA$) corresponds to a maximum velocity width $\approx 400\rm~km~s^{-1}$, which matches most of lines and excludes the continuum. These absorption blocks are regarded as the center for absorption lines. The blocks adjacent to an absorption block have three possibilities, the continuum, the wings of this line or a blended line, and the latter two are also absorption features. One adjacent block will be considered as an extension of the absorption block, if it is not a peak and the difference compared to the higher block next to it is larger than $5\%$ (smaller ones are more likely to be the variation of continuum). Applying Bayesian Blocks to spectra with a S/N of $\approx 20$, one cannot distinguish absorption with depth of $\lesssim 0.05$ unless the length of the block is larger than $20$ resolution elements, which is uncommon. An absorption feature contains an absorption block and its successive extended blocks. Whether these extended blocks are wings or blended lines depends on the symmetry of absorption features, since the wings of a line should be symmetrical. For each absorption feature, we have a central wavelength of the absorption block and two wing widths, which are defined as the distances between the central wavelength and boundaries. If a significant asymmetry is found, this absorption feature will be divided into two absorption lines, where the criterion is that the difference between two wing widths is larger than $0.1\rm ~\AA$ ($2$ bins or $30\rm~km~s^{-1}$, which is the typical $b$ value of a line). This method can identify the case of two blended lines with enough separation, which is also shown in Fig. \ref{f1}.

The use of the Bayesian Blocks algorithm is a significant advantage in our line search methods. This algorithm can identify potential absorption features without determining the local continuum, which avoids the difficulty of assuming a continuum shape in a complex spectrum. Also, this algorithm provides a uniform criterion for line detection, which is controlled by a prior parameter. This parameter is a prior guess for the number of change points, which can be regarded as an indicator of the detection significance. Here, this parameter is fixed to $4$ in our program, approximately equivalent to $4\sigma$. However, our program is also limited by the S/N ratio of spectra, which means weak lines will not be identified, and complex lines will not be adequately decomposed.

\subsection{The Identification of Absorption Systems}
An absorption system should contain several related ions, so we employed a line template match method to find possible systems. Two templates have been considered in this work, one for the high ionization state including two doublets ({\MgX} and {\NeVIII}) and lines from oxygen ({\OIV} and {\OV}) and neon ({\NeIV}, {\NeV} and {\NeVI}). Another template is for low ionization metal lines and galactic lines including strong metal lines within $\lambda \sim 1000-1700~\rm \AA$, which are common in UV spectra \citep{Werk:2013aa, Burchett:2015aa}. These two templates do not overlap since the higher one is $\sim 600 - 800\rm~\AA$, while the lower one is $\lambda \sim 1000-1700~\rm \AA$.

To carry out the line match, one line is considered as the reference line, which must exist in an absorption system. We choose the hydrogen Lyman series lines (mainly Ly$\alpha$ and Ly$\beta$ for different redshift regions) and {\NeVIII} $\lambda\lambda 770.409$ for the two line templates, respectively, since these choices can cover almost all of the redshift region. After a reference line is fixed, the wavelength ratio between two lines is used to judge whether another line is matched. The observed wavelength ratio is a range since observed lines have widths, which are derived from Bayesian Blocks, so if the exact ratio is in this range, two lines are matched. With the most matched lines, a redshift will be marked as a possible absorption system, and the identified lines in this system are excluded from the observed line list. This process is iterated to find all possible absorber systems with more than three spectral lines. Here, three lines is the minimum for possible high ionization state systems: the stronger {\MgX} line, the stronger {\NeVIII} line, and one intermediate line (e.g {\OV} or {\OIV}).

Using this method, false absorption systems can occur by chance because of the large number of lines in each spectrum. The possibility to match a random line is around $1/6$, assuming there are $300$ lines of width $0.3 \rm ~\AA$ in one spectrum, which means the line coverage is $100\rm~\AA$ out of the total wavelength coverage of $650\rm~\AA$. Thus, a three line criterion could have around $100$ false absorption systems by chance, while for systems with five lines, there are only around $10$ false systems. To reduce the number of false systems, we bring in additional information, such as the relative line strengths ($f$ values and abundances) that could physically exist. Among the possible absorption systems, the one at $z = 1.1912$ in LBQS 1435-0134 with $15$ lines is confirmed, and we will introduce this system in Section \ref{LBQS1435}.

\subsection{Absorption Line Measurements}
To obtain line measurements, local continua are obtained by spline fitting in segments of width $\sim 1100 - 3000~\rm km~s^{-1}$, located about the line center. Typically, two sides with width $\approx 300 \rm ~ km~s^{-1}$ are used to do the continuum fitting and to normalize the flux, which ensures that the central $\approx 500\rm~km~s^{-1}$ is left for the line. The wavelength range chosen for continua can be changed by hand if necessary (e.g. when multiple lines appear in one segment, the connection region between two lines will also be included in continuum). With spectra that are normalized by the continuum, the Voigt profile convolved with the COS line spread function (LSF) is used to extract the column density and the $b$ value for the absorption lines in COS/FUV spectra. The fitting is based on the minimum $\chi^2$ method which is realized using the Levenberg-Marquardt optimization algorithm.

During the fitting, we assumed that each absorption system only has one component, so if one absorption system has multiple components, it will be divided into several systems with different redshifts (e.g. the high ionization system in LBQS 1435-0134 has been divided in $1.1910$ and $1.1912$ components). Although we fixed a redshift for each system, velocity shifts were still varied in our fitting because of the uncertainty of redshifts and instrumental shifts \citep{Tumlinson:2013aa}. Thus, for each component, three parameters were varied, including the velocity shift $v$, the velocity factor $b$ value and the column density $N$. For ions in the {\MgX} system, all three parameters are used to fit the profile. If one ion has multiple lines (e.g. the doublet from {\MgX}), these lines will be fit simultaneously with the same parameters. For contaminating lines from other absorption systems, we fixed its shape properties ($b$ and $N$), if they could be obtained by fitting other lines from the same ions. Otherwise, all three parameters are varied to fit a line when a contaminating line is unidentified or there is no way to constrain $b$ and $N$ (e.g. the only line from {\CIII}). For example, the {\OIV} $\lambda\lambda 608.5\rm~\AA$ line of the high ionization system is blended with three Lyman series lines and one {\OIV} line from other systems. These {\HI} lines are fixed, since there are other Lyman lines in this absorption system, while the contaminating {\OIV} is varied because other {\OIV} lines in this system are out of the COS wavelength coverage.

For lines in STIS spectra, besides the profile fitting, we used the apparent optical depth method (AODM; \citealt{Savage:1991aa}) and the curve of growth method (COG) to measure the column density because of the low S/N ratio. For LBQS 1435-0134 with a $21\rm~ksec$ exposure, the S/N ratio of the STIS spectrum is only $\sim 6$ per resolution element, so profile fitting might not work well. However, with the sharp LSF, the AODM should work well for STIS spectra. Both AODM and COG are employed together to determine the consistency of results.

\begin{figure*}
\begin{center}
\includegraphics[width=0.95\textwidth]{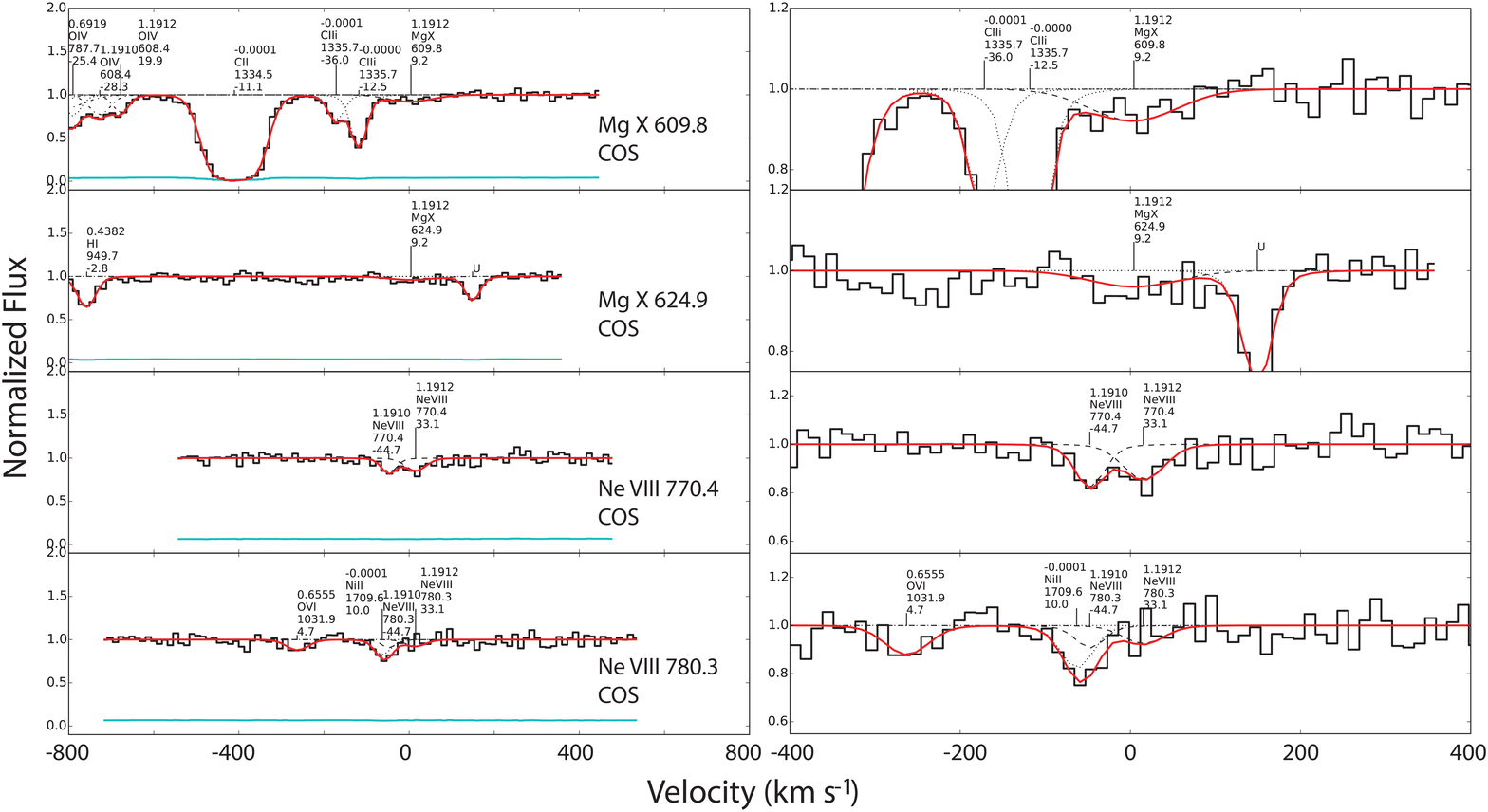}
\end{center}
\caption{Lines due to high ionization species (i.e. {\MgX} and {\NeVIII}) and their zoom-in plots are plotted on the right. The data are the solid black lines and the models are red, while dashed, dotted and dash-dot lines indicate lines in the {\MgX} system and related lines (see the text for details). Every absorption line is marked with its redshift, ion, rest wavelength and velocity shift, while the marker `U' designates unknown lines.}
\label{f2}
\end{figure*}

\begin{table*}
\begin{center}
\caption{Profile fitting results}
\label{fitting_results}
\begin{tabular}{lcccccccccc}
\tableline
\tableline
Ion & $v$ & $b$ & ${\rm log}N$ & $v$ & $b$ & ${\rm log}N$ & ${\rm log} N_{\rm tot}$ \\
 & $\rm km~s^{-1}$ & $\rm km~s^{-1}$ & $\rm cm^{-2}$ & $\rm km~s^{-1}$ & $\rm km~s^{-1}$ & $\rm cm^{-2}$ & $\rm cm^{-2}$ \\
%\tableline
%\multicolumn{8}{c}{ COS profile fitting}\\
\tableline
   {\MgX}\tablenotemark{a}  &                    &                   &                   & $ 9.2\pm 28.0$ &  $68.5\pm 20.3$   &  $13.89\pm 0.10$  & $13.89\pm 0.10$  \\
{\NeVIII}  & $ -104.7\pm 16.7$ &  $20.1\pm 11.5$   &  $13.64\pm 0.17$  & $ 33.1\pm 23.1$ &  $29.9\pm 16.6$   &  $13.67\pm 0.17$  & $13.96\pm 0.17$  \\
  {\NeVI}  & $ -78.4\pm 15.8$ &  $26.4\pm 10.5$   &  $13.98\pm 0.12$  & $ 57.0\pm 29.5$ &  $25.3\pm 19.6$   &  $13.68\pm 0.24$  & $14.16\pm 0.16$  \\
   {\NeV}\tablenotemark{a}  &                    &                   &                   & $-15.2$\tablenotemark{b} &  $46.8\pm 9.2$   &  $13.92\pm 0.06$  & $13.92\pm 0.06$  \\
  {\NeIV}  & $-124.4$\tablenotemark{b} &  $45.3\pm 38.4$   &  $13.44\pm 0.26$  & $3.6$\tablenotemark{b} & $14.6\pm 12.1$  &  $13.29\pm 0.24$  & $13.67\pm 0.25$  \\
  {\OVI}   & $-93.0\pm 2.7$ & $15.7\pm 1.6$ & $14.39\pm 0.03$ & $10.8\pm3.0$ & $17.2\pm 2.1$ & $14.10\pm 0.04$ & $14.57\pm 0.03$\\
    {\OV}  & $ -104.8\pm  3.5$ &  $18.2\pm  2.0$   &  $14.07\pm 0.04$  & $ 4.8\pm  4.4$ &  $22.2\pm  2.3$   &  $13.98\pm 0.04$  & $14.33\pm 0.04$  \\
   {\OIV}  & $ -97.5\pm 14.9$ &  $27.9\pm  7.8$   &  $14.17\pm 0.12$  & $ 6.4\pm 12.9$ &  $21.0\pm  6.1$   &  $14.03\pm 0.15$  & $14.41\pm 0.13$  \\
  {\OIII}\tablenotemark{a}  &                    &                   &                   & $-22.8\pm 35.8$ &  $52.2\pm 24.7$   &  $13.45\pm 0.15$  & $13.45\pm 0.15$  \\
   {\NIV}  & $-120.0\pm 24.54$ &  $22.7\pm 19.2$   &  $12.68\pm 0.22$  & $  -7.4\pm 50.1$ &  $ 9.0\pm 23.6$   &  $12.23\pm 0.53$  & $12.81_{-0.28}^{+0.32}$  \\
   {\HI} & $-87.75\pm13.85$ & $35.5 \pm 6.8$ & $ 13.85 \pm 0.09$ & $28.28\pm39.91$ & $33.2 \pm 17.3$ & $13.30   \pm 0.31$ & $13.97\pm 0.13$\\
\tableline
\end{tabular}
\tablenotetext{1}{For these three ions, we only fit them using one component model, since we cannot distinguish two components resulted by the low S/N ratio.}
\tablenotetext{2}{For {\NeV} and {\NeIV}, we fixed $v$ to limit the error bar. The fixed $v$ is the best fitting value when it is varied.}
\end{center}
\end{table*}

\section{LBQS 1435-0134}
\label{LBQS1435}
\subsection{Absorption Systems}
\label{absorbers}
Nineteen absorption systems have been identified in the LBQS 1435-0134 spectra using our method. All of these absorption systems are identified using the strategy stated in Section \ref{method}, and eighteen systems are based on the low ionization line template, while the only one at $z = 1.1912$ is from the high ionization line template. More than 250 lines (including blended lines) have been identified, which do not include isolated Ly$\alpha$ systems. A full summary of the lines are beyond the scope of this paper and will be reported elsewhere.

\begin{figure*}
\begin{center}
\includegraphics[width=0.95\textwidth]{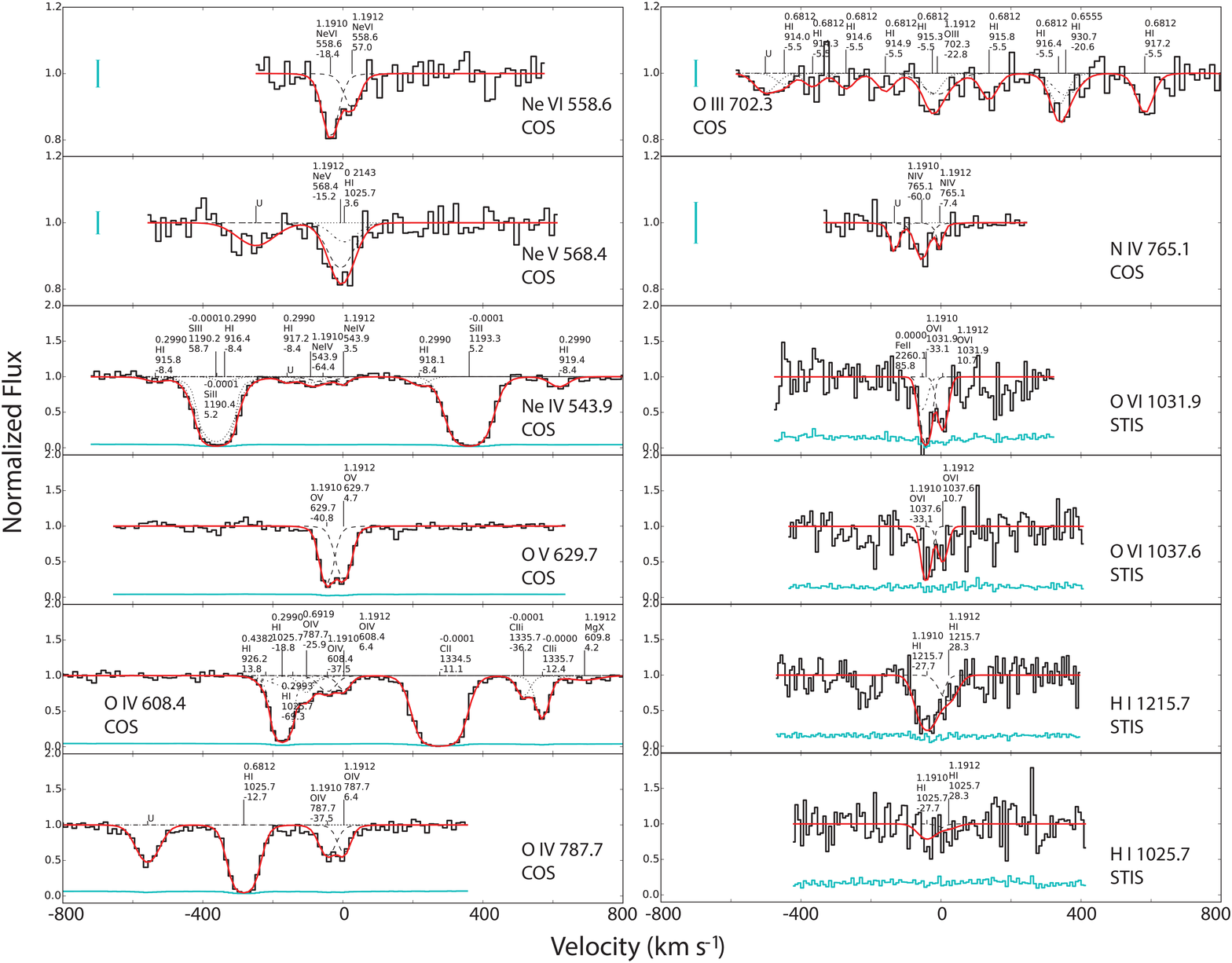}
\end{center}
\caption{Other lines in the $z=1.1912$ system are shown except for the two {\OIV} blended with non-redshifted {\HI} features. Symbols in this figure are the same as in Fig. \ref{f2}, except for Ly$\beta$, where the red line is the projected shape of Ly$\alpha$. In the zoom-in plots, characteristic errorbars are given as cyan markers.}
\label{f3}
\end{figure*}

\subsection{The {\MgX} System}
\label{high_system}
For the high ionization absorption system at $z = 1.1912$, there are $17$ lines from $11$ ions in the wavelength coverage. Fourteen lines lie in the COS spectrum and three lines lie in the STIS spectrum. Two lines from {\OIV} $\lambda \lambda~ 553.329~ \rm \AA$ and $554.076~ \rm \AA$ are blended with galactic Ly$\alpha$ absorption and the geocoronal hydrogen emission, which is not distinguishable. Using the profile fitting method stated in Section \ref{method}, we fit all lines in the COS spectrum except for the two {\OIV} lines. The fitted lines are shown in Fig. \ref{f2} and Fig. \ref{f3} for high ionization lines and other low or intermediate ionization lines, respectively. The results of the column density and the $b$ value are summarized in Table \ref{fitting_results}. The velocity shifts given in the table are aligned to $z = 1.1912$ with a typical separation of $110-140 \kms$ between the two components. We note that the our single component fitting on {\MgX} may not represent the true structure of this ion, which would only be revealed by higher S/N data. According to the position and the width of the {\MgX} doublet, we suggest that this ion has a similar shape of {\NeVIII}. We also tried a two component model to the {\MgX} doublet, which yielded a total column density of ${\rm log}N = 13.95_{-0.20}^{+0.21}$. This column density is consistent with the single component model because this line is in the low optical depth regime, where the column is nearly independent of the line shape.

{\MgX} is the crucial ion in our study, but the lines from it are weak because of the low abundance and the small maximum ionization fraction ($\approx 22\%$). As shown in Fig \ref{f2}, the depth of the stronger {\MgX} line is $\approx 10\%$, while the weaker line is $\approx 5\%$. To determine the significance of the doublets from {\MgX}, we fit these two lines with and without {\MgX}. Between the two fittings, the change of $\chi^2$ is $43.0$ from $225.2$ to $268.2$, which can be separated into $28.7$ (from $126.6$ to $155.3$) for the stronger line and $14.3$ (from $98.6$ to $112.9$) for the weaker line. Considering the change of the degree of freedom (dof) is $3$, the total significance of the doublets is $5.8~\sigma$, with $4.6~\sigma$ for the stronger line and $2.8~\sigma$ for the weaker line. Using the same method, {\NeVIII} shows a total significance larger than $6.5~\sigma$, which confirms its existence. We also calculate the significance level of each line using the method described in \citet{Keeney:2012aa}, and the results are consistent with the $\chi^2$ method, showing $4.6\sigma$, $2.7\sigma$ and $6.8\sigma$ for the {\MgX} doublet lines and the strong {\NeVIII} line, with EW of $38.6\pm 7.0~\rm m\AA$, $22.8\pm7.0~\rm m\AA$ and $82.4\pm 11.8~\rm m\AA$, respectively. The weaker {\NeVIII} line is blended with a galactic {\NiII} line $\lambda\lambda1709.600~\rm\AA$, so the equivalent width is poorly constrained.

In the STIS spectrum, both the {\OVI} doublet and Ly$\alpha$ are matched at the expected positions. For the {\OVI} doublet, the stronger line is blended with another line, which should be the galactic {\FeII} $\lambda\lambda 2260.079$ line. The weaker line is used to obtain the line strength, showing ${\rm log}N = 14.49\pm 0.06$ in AODM and $14.50\pm0.10$ in COG (equivalent width, EW $= 295\pm 33\rm~m\AA$; $b \approx 20 \rm~ km~s^{-1}$, which is consistent with {\NeVI} and {\OV}). In the profile fitting, we obtain a total {\OVI} column density of $14.57\pm 0.03$, which may be affected by the galactic {\FeII}. For the Lyman series, only Ly$\alpha$ has been detected, so just one line is used, showing ${\rm log}N = 13.93\pm 0.04$ using the AODM approach. With the COG approach, since the $b$ value affects the EW significantly in the saturated region, we assume the $b$ value is $\approx 30 - 40~\rm km~s^{-1}$ because of the total velocity width of $\approx 160~\rm km~ s^{-1}$. Thus, the derived column density is ${\rm log}N = 13.99\pm0.13 \rm~ cm^{-2}$ for an EW $= 607\pm45\rm~m\AA$ in COG. The profile fitting of Ly$\alpha$ shows a consistent column density of ${\rm log}N = 13.97\pm 0.13$. This column density is also consistent with the constraint of the weak Ly$\beta$ line. The expected equivalent width of Ly$\beta$ is $\approx 141\rm~m\AA$, which is consistent with the measured equivalent width of $146\pm 54\rm~m\AA$, a detection slight below $3\sigma$. For these two ions, we adopt the values from AODM in the following analysis.

\begin{figure*}
\begin{center}
\plottwo{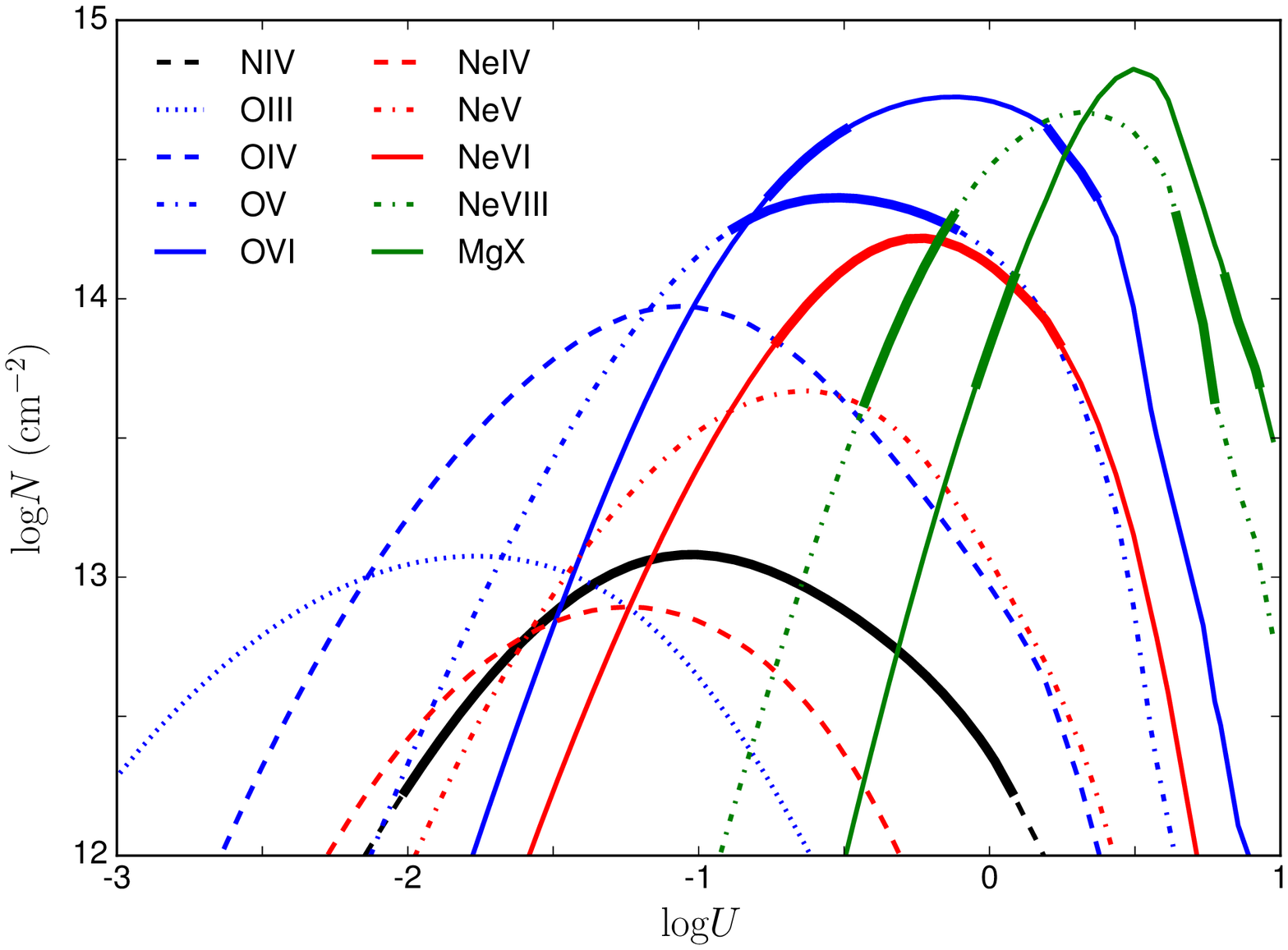}{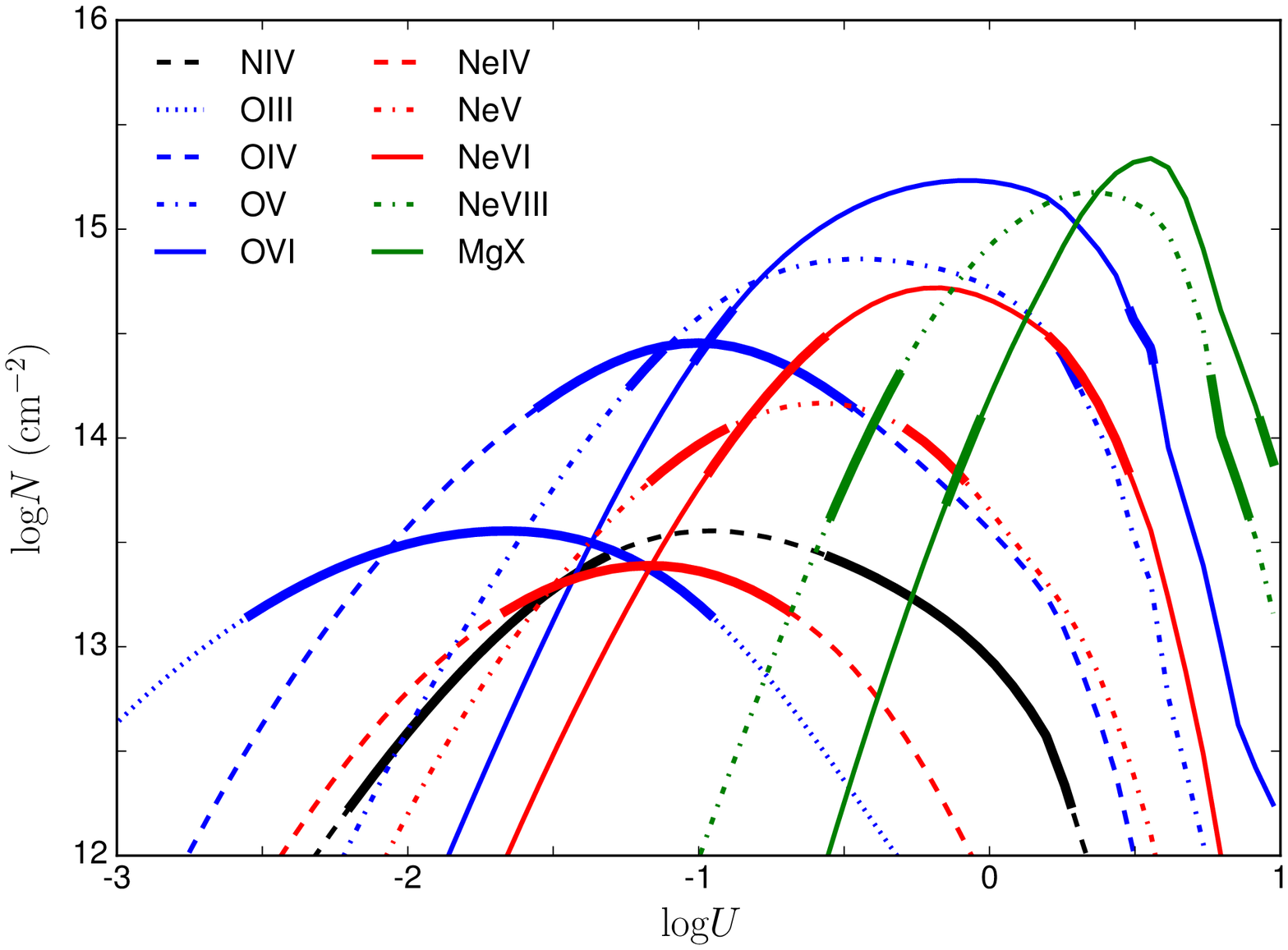}
\end{center}
\caption{The left panel shows the photoionization model with $[Z/X] = -0.5$, while the right panel is $[Z/X] = 0$. The bold solid lines indicate the observational measurements with $2\sigma$ error bars.}
\label{f4}
\end{figure*}

After measuring lines from ground levels, we also examined the spectra for the presence of density sensitive lines from excited levels (e.g. {\OIV}$^{*}$ and {\OV}$^{*}$) in the COS spectrum. A fairly strong line shows up at the position of {\OV}$^*$ $\lambda \lambda ~759.442$. However, we rule out the identification of this line as redshifted {\OV}$^*$ $\lambda \lambda ~759.442$ for three reasons. First, this line is also identified as {\OVI} $\lambda \lambda~ 1031.912$ at $z=0.6127$ and this doublet can be fitted well without an additional line. Secondly, {\OV}$^*$ has two strong lines ($\lambda\lambda~759.442$ and $\lambda\lambda~ 760.446$) with similar $f$ values ($0.1913$ and $0.1432$) and similar low excited levels ($10.16$ and $10.21\rm~eV$ with critical density of $\sim 10^5\rm~cm^{-3}$), however, no second line occurs at the expected position. Thus, the absence of another {\OV}$^*$ line does not support the existence of density sensitive lines. Also, the {\OIV}$^*$ line is absent. {\OIV} has a similar column density to {\OV} in this system and {\OIV}$^*$ lines require a much smaller excited energy ($0.05~\rm eV$) requiring a lower critical density ($\sim 10^3\rm~cm^{-3}$). For these three reasons, no density sensitive line has been detected in this high ionization state system.

\begin{figure*}
\begin{center}
\plottwo{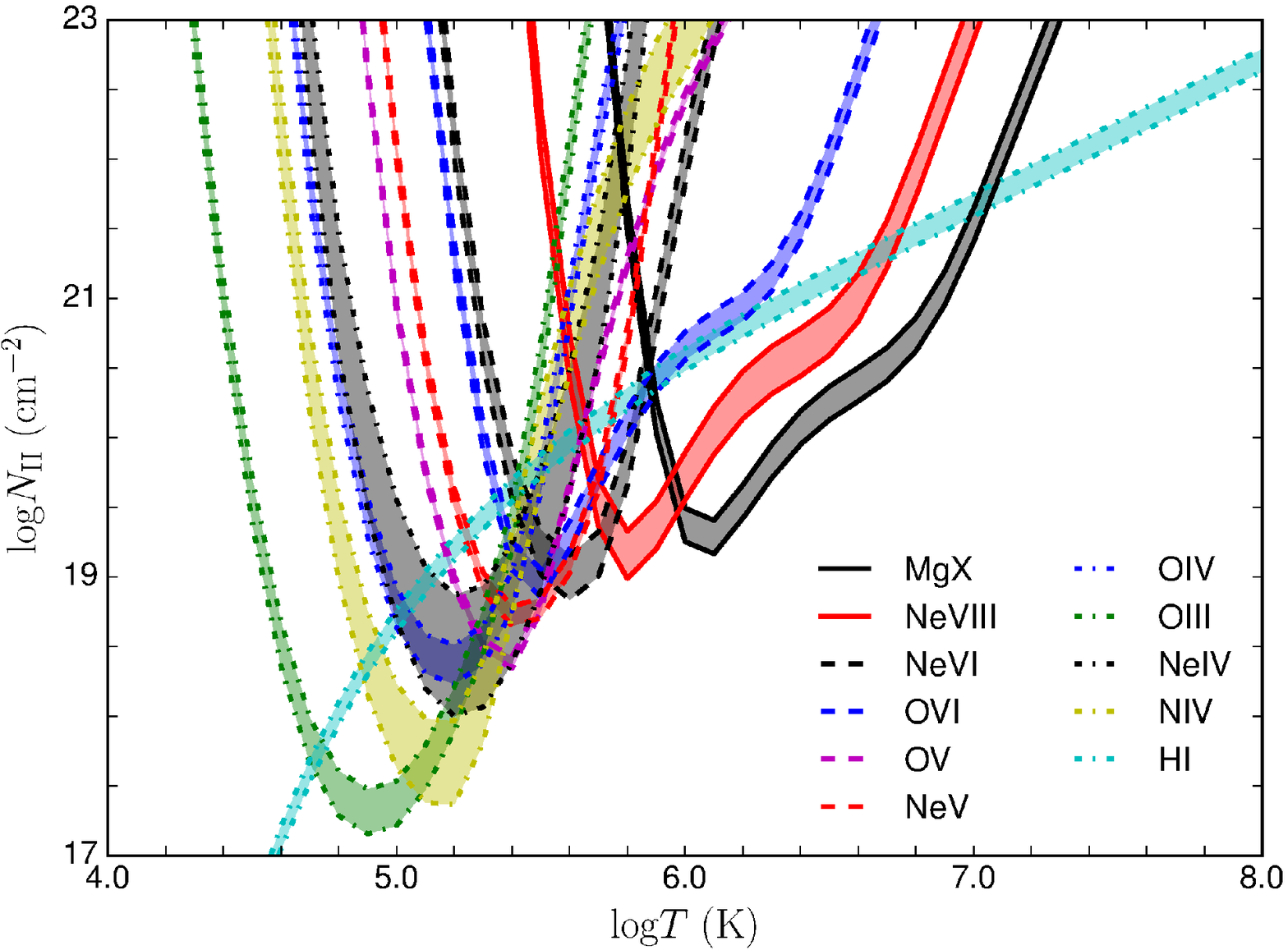}{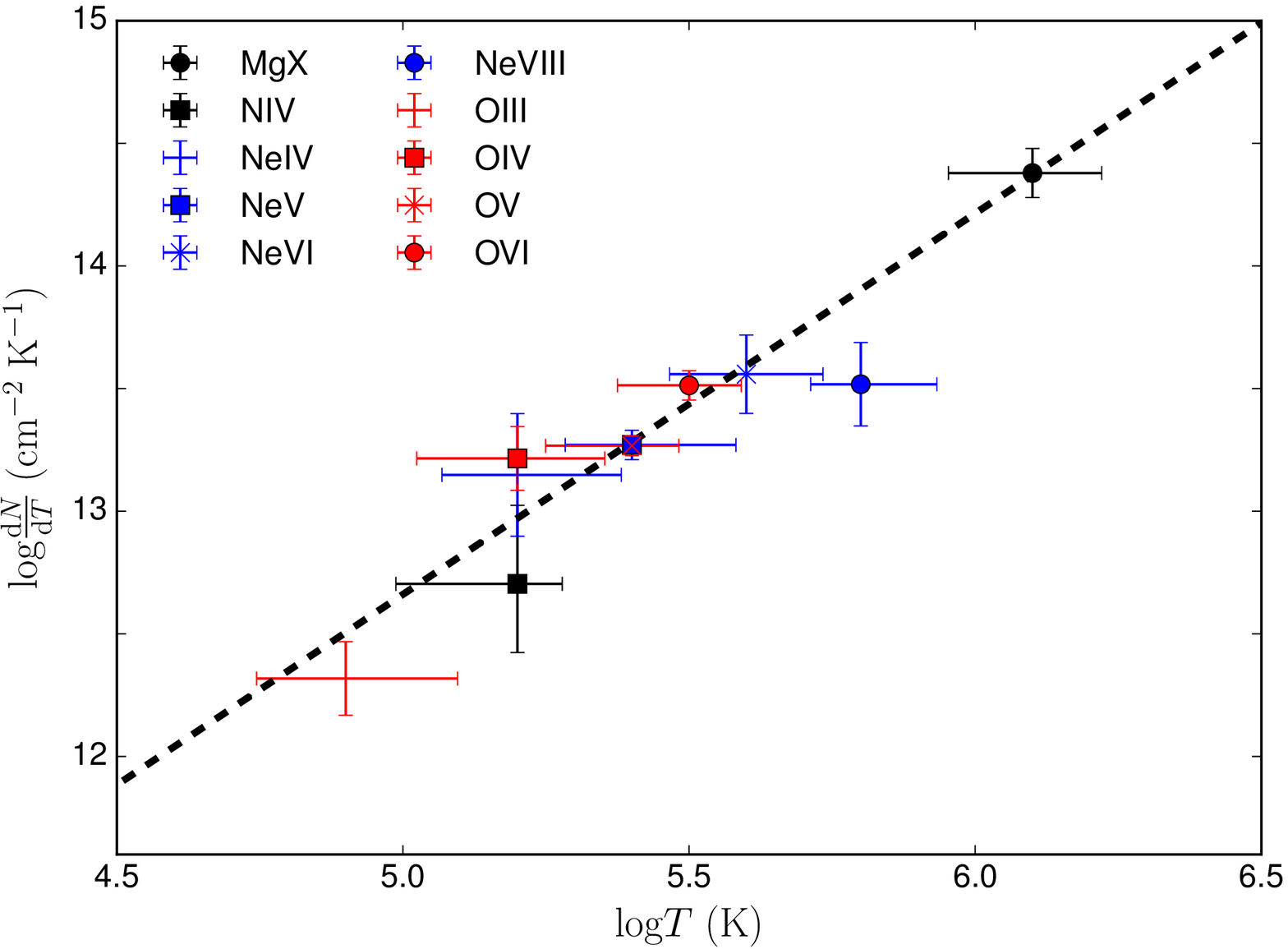}
\end{center}
\caption{The left panel is the $T-N$ diagram for the collisional ionization equilibrium model. For each ion, the band indicates the possible region for its column density. The right panel shows the fitting results of the power law model. The dashed line shows the temperature dependence of the hydrogen column density. For each ion, the temperature is the peak temperature of the ionization fraction, and the error bar is the full width of the half maximum, while the gradient of the column density is the modified gradient of hydrogen $\frac{{\rm d}N_{\rm H, ion}}{{\rm d}T} = \frac{{\rm d}N_{\rm H, model}}{{\rm d}T} \times N_{\rm ion, observed} / N_{\rm ion, model}$.}
\label{f5}
\end{figure*}

\section{Model}
\label{model}
For the analysis of the {\MgX} system, two models, the photoionization model and the collisional ionization model, are considered to model the ionization states in the gas. Here, we only show the details in the application of ionization models, and the nature of this gas will be discussed in Section \ref{PIorCI}. Also, the combination of the photoionization and the collisional ionization model is beyond the scope of this paper, so we deal with these two models separately.

\subsection{The Photoionization Model}
\label{PI}
We employed the photoionization code CLOUDY (version 13.03; \citealt{Ferland:1998aa}), which can calculate the column density for each ion with the input radiation field and the absorption gas. The input radiation field is the HM05 in CLOUDY, which is the cumulative background radiation field from AGN as calculated by \citet{Haardt:1996aa} and updated by Haardt \& Madau (2005 private communication to CLOUDY). This integrated radiation field is believed to be redshift dependent and uniform at a given redshift. Here, we fixed the redshift to $1.19$, and assumed the absorption system is plane-parallel to simplify the calculation. When carrying out the simulation, we fixed the neutral hydrogen column density to the observed value as the boundary condition, and varied the density to achieve different ionization parameters $U = n_{\gamma}/n_{\rm H}$, where $n_{\gamma}$ is the photon density with energies higher than $13.6\rm~ eV$ and $n_{\rm H}$ is the hydrogen density. 

In Fig. \ref{f4}, we show two models for the metallicity of $[Z/X]=-0.5$ and $[Z/X]=0$, where $[Z/X] = {\rm log}(Z/X)-{\rm log}(Z/X)_{\odot}$, and the used solar metallicity is from \citet{Asplund:2009aa}. In these two plots,  the column density dependence on the ionization parameter are plotted for all detected ions except for {\HI}, which is fixed to $10^{13.93}\rm~cm^{-2}$, and the observed column densities with $2 \sigma$ error bars, shown as bold lines for each ion. We note that the lower observed limits are still higher than the peak of {\NeV} and {\OIV} in the $[Z/X]=-0.5$ model, which is odd since these two ions do not have similar ionizational potentials. This is mainly because of the small {\HI} column density (${\rm log}N = 13.93$) compared to the COS-Halo sample ($\gtrsim 15$; \citealt{Werk:2014aa}). The standard approach of fixing this problem is to raise the metallicity to obtain detectable metal lines, so we show the one solar metallicity model.

In the one solar metallicity model, there is no single ionization parameter that reproduces all the observed ions. If we consider a two-components model, intermediate ions through \NeVI admit to a solution at ${\rm log} U \approx -1.3 \pm 0.3$, which corresponds to a hydrogen density of $10^{-3.5}\rm~ cm^{-3}$, a mean temperature of $2.3\times 10^4\rm~ K$, and a path length of $1.3\rm~ kpc$. A second component is required for the two higher ionization species of {\NeVIII} and {\MgX}, where a solution is found at ${\rm log} U \approx 0.9$. This component corresponds to a density of $10^{-5.7}\rm~ cm^{-3}$, a mean temperature of $1.5\times 10^6\rm~ K$, and a path length of $1.1\times 10^3\rm~Mpc$. This unreasonably large path length is a result of requiring the high ionization state also to produce the total observed {\HI} column density, which should be fixed. However, using the current data, there is no way to divide the observed {\HI} column density into two parts for the low and high states, respectively. We also consider the case where the observed {\HI} from the high state is at its minimum to reproduce the observed {\MgX} and {\NeVIII}. In this case, the solution should be around the peak of the {\MgX} and {\NeVIII}, and show an ionization parameter of ${\rm log} U \approx 0.5$, which corresponds to a hydrogen column density of $\sim 10^{19.5}\rm~cm^{-2}$, a density of $\sim 10^{-5.5}\rm~cm^{-3}$, and a path length of $\sim 3~\rm Mpc$. This path length is still about three orders of magnitude larger than the lower ionization state solution. This size difference indicates that they are not part of the same cloud but would represent different phases. Whether such phases are reasonable are explored in Section \ref{PIorCI} after we exclude the possibility of AGN outflow.

\subsection{The Collisional Ionization Model}
\label{CIM}
We consider the collisional ionization model for the {\MgX} gas because of the difficulty of reproducing {\MgX} and {\NeVIII} in the photoionization model. Here, we assume collisional ionization equilibrium (CIE) for this {\MgX} gas, using the calculations of \citet{Bryans:2006aa}. The temperature-hydrogen column density ($T-N$) diagram for observed ions is given in Fig. \ref{f5}. The width of the band for each ion takes into account the uncertainties in the column densities. The relative abundance of metals are set to the solar abundance and the metallicity of $[Z/X]=-0.5$.

The single temperature model assumes ions originate from the same gas with a uniform temperature and a given column density, and one may expect that these ions go across a single point in the $T-N$ diagram. We find that not all of the ions can be reproduced by one single temperature model. If only considering {\MgX} and {\NeVIII}, they can be modeled by a gas with a temperature of $10^{6.00\pm 0.04}\rm~K$ and a hydrogen column density of $10^{19.67\pm 0.16}\rm~cm^{-3}$. However, {\NeVI} and {\OVI} cannot be reproduced with these parameters, so a multiple temperature medium is required even only for the high ionization state.

As shown in Fig. \ref{f5}, multiple intermediate ionization species occur near a temperature of $10^{5.5}\rm~K$ and a hydrogen column density of $10^{19.1}\rm~cm^{-2}$, suggesting that a two-temperature fit might work. A best-fit for a two temperature model yields the parameters ${\rm log} N_1 = 18.60 \pm 0.07$ at ${\rm log} T_1 = 5.33\pm 0.02$ and ${\rm log} N_2 = 20.34 \pm 0.09$ at ${\rm log} T_2 = 5.89 \pm 0.01$. However, this is not an acceptable fit as $\chi^2 = 19.4 (7~ {\rm dof})$, with a significant fraction of the $\chi^2$ coming from {\NeVIII} and {\NeV}. This deviation can not be mitigated by adjusting the relative abundance of neon, since the observed {\NeVIII} column density is much lower than the prediction, while {\NeV} is opposite. Thus, we reject this two temperature model. 

The next level of complexity is a three temperature model, which is more successful. The best fitting parameters are ${\rm log} N_1 = 18.48\pm0.14$ at ${\rm log} T_1 = 5.27\pm0.04$, ${\rm log} N_2 = 19.00\pm0.21$ at ${\rm log} T_2 = 5.59\pm0.07$ and ${\rm log} N_3 = 20.06\pm0.16$ at ${\rm log} T_3 = 5.94\pm0.03$. The resulting $\chi^2$ is $5.5$ and the dof is $5$, an acceptable fit. 

This three temperature model has six free parameters for eleven data points, so we also consider a power-law model with only four free parameters. In this model, we adopt a power law between the temperature and the column density ${\rm log} ({\rm d} N / {\rm d} T) = a + B{\rm log}T$. Along with parameters $a$ and $B$, there are the lower and upper limits of temperature ($T_{\rm down}$ and $T_{\rm up}$). Best-fit results show that $a = 4.3 \pm 2.2$, $B = 1.55 \pm 0.41$, $T_{\rm down} = 10^{4.39\pm0.13}\rm~K$ and $T_{\rm up} = 10^{6.04\pm0.05} \rm~K$ with $\chi^2 = 3.3 (7)$, which is an acceptable fit. As shown in Fig. \ref{f5} and Table \ref{comparison}, the high ionization gas is well modeled by a power law model with the maximum temperature larger than $10^6~\rm K$ and the total hydrogen column density $\approx 8.2\times 10^{19} (0.3 Z_{\odot}/Z) \rm~ cm^{-2}$.

\begin{table}
\begin{center}
\caption{Model comparison}
\label{comparison}
\begin{tabular}{lcccc}
\tableline
\tableline
Ion & ${\rm log}N_{\rm adopted}$ & ${\rm log}N_{2T}$ & ${\rm log}N_{3T}$ & ${\rm log}N_{\rm pl}$\\
\tableline
Mg {\scriptsize X}    & $13.89\pm 0.10$ & $13.78$ & $13.93$ & $13.88$\\
Ne {\scriptsize VIII} & $13.96\pm 0.17$ & $14.95$ & $14.51$ & $14.35$\\
Ne {\scriptsize VI}   & $14.16\pm 0.16$ & $13.72$ & $14.17$ & $14.19$\\
O {\scriptsize VI}    & $14.49\pm 0.06$ & $14.48$ & $14.46$ & $14.41$\\
Ne {\scriptsize V}    & $13.92\pm 0.06$ & $13.66$ & $13.97$ & $13.93$\\
O {\scriptsize V}     & $14.33\pm 0.04$ & $14.47$ & $14.32$ & $14.34$\\
Ne {\scriptsize IV}   & $13.67\pm 0.25$ & $13.69$ & $13.73$ & $13.49$\\
O {\scriptsize IV}    & $14.41\pm 0.13$ & $14.39$ & $14.43$ & $14.17$\\
N {\scriptsize IV}    & $12.81_{-0.28}^{+0.32}$ & $13.06$ & $13.33$ & $13.08$\\
O {\scriptsize III}   & $13.45\pm 0.15$ & $13.16$ & $13.42$ & $13.64$\\
H {\scriptsize I}     & $13.93\pm 0.04$ & $13.91$ & $13.72$ & $13.93$\\
\tableline
\end{tabular}
\end{center}
\end{table}

\section{Discussion}
\label{discussion}
We have examined an absorption system at $z = 1.1912$ in the sightline toward \object{LBQS 1435-0134} with high ionization species (i.e. {\MgX} and {\NeVIII}) and several intermediate or low ionization species including {\OVI} and {\HI}. We consider whether the observed absorption system is likely produced by the extended hot halo of a galaxy and if the gas, especially the {\NeVIII} and {\MgX} ions, is collisionally ionized or photoionized. 

\subsection{An Intrinsic AGN Outflow?}
\label{AGN_D}
Before modeling the high ionization state system, we consider whether the {\MgX} is due to the AGN outflow. As introduced in Section \ref{intro}, previous observations show that {\MgX} could be common in the intrinsic absorption of AGN rather than absorption by the hot halo of a foreground galaxy. A hot gaseous galaxy halo cannot be confirmed before its host galaxy is discovered, however, with the current observation and the guidance from previous studies, it is possible to show that absorption by a hot halo is the more likely explanation.

Some AGNs can have very high velocity outflows (up to $\sim 0.2~ c$), while the velocity widths can be quite narrow ($\sim  100 \rm ~ km ~ s^{-1}$), which may be mistaken as intervening absorption systems \citep{Misawa:2007aa}. These high velocity outflows sometimes can be distinguished from intervening systems, if they show absorption line variability, partial covering, or a density sensitive line from excited states \citep{Teng:2013aa, Muzahid:2013aa, Finn:2014aa, Muzahid:2016ab}. It is believed that absorption systems with high velocities relative to the AGN are due to the wind originating from the disk, which implies that the variability should be common among the AGN intrinsic absorption; this variability has been observed \citep{Teng:2013aa, Muzahid:2016ab}. The signature of partial covering is that the absorption line is saturated but not black. The phenomenon indicates the length scale of the gas is comparable to the disk of an AGN ($\sim 10\rm~pc$; \citealt{Sulentic:2000aa}), and it is about four order of magnitude lower than the virial radius of galaxies (several hundreds $\rm kpc$). The existence of absorption from excited states indicates large densities ($\gtrsim 10^2\rm~cm^{-3}$ for {\OIV}$^*$), that would not occur in the galaxy halo (several orders of magnitude larger than $n_{200}$). Thus, these diagnostics can be employed to confirm AGN outflows, although there can still be intrinsic AGN absorption systems that do not have these features.
 
Phenomenologically, we want to compare this {\MgX} system to the observed {\MgX} AGN outflows, so we give a brief summary for the current results on {\MgX} AGN outflows. There are nine {\MgX} outflows in five AGN reported in the literature \citep{Muzahid:2012aa, Muzahid:2013aa, Finn:2014aa}. Eight of nine absorption systems have relatively large {\MgX} column densities ($\gtrsim 10^{14.7}\rm~cm^{-2}$), which indicates large total column densities of $\gtrsim 10^{21}\rm~cm^{-2}$. However, with the common existence of {\OIV}$^*$ lines (four of five objects), the typical path length is several parsecs, which is consistent with the partial covering model (\citealt{Arav:2013aa, Finn:2014aa}; we checked the {\OIV}$^*$ for PG 1206+459 and PG 1338+416). Also, these outflows can be modeled by the multi-phase photoionized gas models \citep{Arav:2013aa}. The {\MgX} system in LBQS 1435-0134 seems to be significantly different from these {\MgX} AGN outflows with the smaller column density and the lower density. Thus, although we cannot exclude the possibility of AGN outflow completely, this {\MgX} system is unlikely to be an AGN outflow.

\subsection{Photoionization or Collisional Ionization?}
\label{PIorCI}
For intervening absorption systems, previous studies show that a photoionization model with ${\rm log} U \sim -5$ to $-1$ can be used for the low ionization species, while the origin of the higher ionization species is more complicated. About half of {\OVI} and all of {\NeVIII} intervening systems are difficult to be reproduced in the photoionization model, which indicates they should be collisionally ionized in the high temperature medium \citep{Thom:2008aa, Narayanan:2012aa, Werk:2014aa}. Here, we will show that {\MgX} is also a tracer for high temperature medium (even higher than $10^6\rm~ K$).

As stated in Section \ref{PI}, there should be two components in the one solar metallicity model to account for intermediate and high ionization species in the photoionization model. The component for intermediate and low species has a ionization parameter of ${\rm log} U \approx -1.3$. With this ionization parameter, the required density ($10^{-3.5}\rm~cm^{-3}$) and the total column density ($10^{18}\rm~cm^{-2}$) are consistent with the COS-Halo sample (densities of $10^{-1}$ to $10^{-4}~\rm cm^{-3}$ and hydrogen column densities of $10^{17}$ to $10^{20}~\rm cm^{-2}$; \citealt{Werk:2014aa}). The path length of $1.3\rm~ kpc$ indicates this component could be a cool cloud in the halo. However, the high ionization state with ${\rm log} U \sim 0.9$ has a path length of $1100~\rm Mpc$. This extremely large path length leads to a line broadening due to Hubble flow that is about half the speed of light, which is opposite to the observed narrow line ($\lesssim 100\rm~km~s^{-1}$). Even if we consider the modification of the {\HI} column density, the low limit of the ionization parameter is $0.5$, which results a density of $3\times 10^{-6}\cc$ (more than one order of magnitude lower than $n_{200}$), a path length of $\approx 3\rm~ Mpc$ and a Hubble flow of $200\rm~km~s^{-1}$ (twice of the observed value). Therefore, the photoionization solution for the intermediate and low species seems to be plausible, while the one for the high ionization species is unreasonable.

The collisional ionization model is required to model the high ionization species {\MgX} and {\NeVIII}. As shown in Fig. \ref{f5}, the possible single temperature solution is around the temperature of $10^{6.0}\rm~K$, which indicates the existence of the hot gas. With the corresponding total column density of $\sim 10^{19.9}\rm~cm^{-2}$, the density of this absorption gas can not be lower than $2.9\times 10^{-5}\rm~cm^{-3}$ to avoid the path length larger than $1\rm~Mpc$, which will result into an unreasonable width due to the Hubble flow. Then, we adopt the density of $5\times 10^{-5}\rm~cm^{-3}$ ($n_{200}$) which leads to a thermal pressure of $50\rm~ K~ cm^{-3}$.

It is of interest to show whether the intermediate and low ionization gas can be in pressure equilibrium with this hot gas in the photoionization model or the collisional ionization model. In the photoionization model, for the intermediate ionization state species, the predicted temperature is $2.3\times 10^4\rm~K$, and the density is $3.1\times 10^{-4}\rm~cm^{-3}$, which implies a pressure of $7\rm~K~ cm^{-3}$. This pressure is one order of magnitude lower than the hot gas inferred from the collisional ionization model, which indicates the possible photoionized gas cannot be in pressure equilibrium with the hot gas. In the collisional ionization model, the variation of the temperature is about one order of magnitude, so to make the gas have uniform pressure, the density should also have a variation of one order of magnitude. With the lower limit of $2.9\times 10^{-5}\rm~cm^{-3}$, the density could be up to $1.6\times 10^{-3}\rm~cm^{-3}$. Therefore, we suggest that even for the intermediate and low ionization gas, the photoionization is unlikely to be the only origin of the ionization species due to the difficulty to build the pressure balance.

For the collisional ionization model, the Doppler $b$ value can constrain the nature of the gas to analyze the turbulence or other broadening mechanisms. In the power law model, we assume that for each ion except for {\HI}, the temperature is approximately given by the peak of the ionization fraction, which also accounts for most of the column density of that ion. Because of the lower limit of the temperature, approximately $80\%$ of {\HI} is in the temperature region between $10^{4.3}$ to $10^{5.7} \rm~ K$, so we adopt the temperature of ${\rm log} T \approx 5.0$ as the typical temperature. Based on these temperature constraints, thermal limits of $b$ values for metal ions are constrained to be in the range $20 - 30\rm~km~s^{-1}$, while it is $40\rm~km~s^{-1}$ for {\HI}. For ions that can be resolved into two components, this range is consistent with $b$ values shown in Table \ref{fitting_results}, while {\OVI} and {\HI} are consistent with the assumption we used to measure the column density. This consistency indicates that most of ions are in a quiescent state without significant turbulence or convection.

\subsection{Interaction Layers or A Hot Halo?}
\label{InteractionOrHalo}
There are two possibilities for the intervening absorption due to foreground galaxies -- the interaction layer or the hot halo. The interaction layer is the interface between a cool cloud ($\sim 10^3 \rm~ K$) and the surrounding hot gas ($\sim 10^6 \rm~ K$) in the halo \citep{Kwak:2010aa}. In this mixing interface, multiple intermediate ions can be produced, such as {\OVI}, {\OV} and {\SiIV}. Based on our modeling results, we may distinguish these two origins for the {\MgX} system.

Previous works on intervening {\NeVIII} systems argued that high ionization state ions may also be produced in this interaction layer because of their relatively lower temperature ${\rm log} T \sim 5.6-5.7$ \citep{Narayanan:2011aa, Savage:2011ab, Narayanan:2012aa}. Here, they used the single temperature model to reproduce {\NeVIII} and {\OVI}. However, the assumption required in this model is not substantiated. To apply the single temperature model, it is assumed that {\NeVIII} and {\OVI} are from the same gas and both collisionally ionized. Nevertheless, previous studies on intervening {\OVI} surveys show that it could be either collisionally ionized or photoionized \citep{Tripp:2008aa, Thom:2008aa}, which means using {\OVI} to constrain the collisionally ionization model could result in an incorrect result. Even if {\OVI} is collisionally ionized, recent studies on the Milky Way hot halo show that the hydrogen column density required to reproduce {\OVI} is about one order of magnitude lower than the one derived from {\OVII} (a higher temperature medium), which implies {\OVI} is not cospatial with the higher temperature gas \citep{Hodges-Kluck:2016aa}. With an ionizational potential of $207.3\rm~eV$, {\NeVIII} traces a higher temperature gas, which is not cospatial with the {\OVI} gas, so the combination of {\NeVIII} and {\OVI} will result in a lower temperature. {\MgX} has a peak ionization fraction at $10^{6.1} \rm~ K$ and has a full width of half maximum of $7.9\times10^5 \rm~K$. Therefore, this ion occurs at the virial temperature of a $L\sim 0.2-2~L^*$ galaxy, which is also the temperature of the hot medium in interaction layer models. Thus, we suggest the gas traced by {\MgX} and {\NeVIII} is the hot halo rather than the interaction layer.

Another consideration is the intermediate ionization species column density in the interaction layer model. Currently, there are two theoretical models focusing on the interaction layer, the turbulence mixing layer model and the conductively evaporating model \citep{Kwak:2010aa, Gnat:2010aa, Kwak:2015aa}. The turbulence mixing layer model predicts $10^{13}$ to $10^{14} \rm~cm^{-2}$ {\OVI} for one cloud, which means one expects around $10$ clouds to reproduce the observed {\OVI} column density. However, the expected number of cool cloud in one sightline is about $2$, which can be estimated by the coverage rate ($\approx 90\%$) in the COS-Halo's sample assuming the Poisson distribution  \citep{Werk:2014aa}. The conductively evaporating model even predicts a lower {\OVI} column density ($10^{12.5}\rm~cm^{-2}$ per layer) than the turbulence mixing layer model, which results in a larger gap between the expected cool cloud number and the required one. Therefore, this significant gap also suggests that it is unlikely for all of the observed {\OVI} column density to be due to interaction layers. Also, current mixing layer calculations do not include the columns of {\NeVIII} or {\MgX}, which makes it still a question that whether interaction layers can produce detectable high ionization species.

One more consideration is on the Doppler $b$ values of metal lines, which reflect the dynamical information of the gas. As stated in Section \ref{PIorCI}, the gas is quiescent, showing the $b$ value at the thermal velocity limit, which is opposite to the expectation of interaction layer models \citep{Gnat:2010aa, Kwak:2015aa}. In either turbulence mixing layer model or conductively evaporating model, turbulence or convection is inevitable because of the low efficiency of radiation and thermal conduction due to the low density. The turbulence velocity is about $100\rm~km~s^{-1}$ \citep{Kwak:2010aa}, so small $b$ values at thermal limits ($\sim 20-40\rm~km~s^{-2}$) imply a volume-filled quiescent halo rather an active interaction layer. Also, small $b$ values in our case are not unique for high ionization states in observations, which have been confirmed to be associated with galaxies. \citet{Narayanan:2011aa} and \citet{Meiring:2013aa} showed both {\OVI} and {\NeVIII} in PKS 0405-123 and PG 1148+549 have small $b$ values around $20-30 \rm~km~s^{-1}$. Based on these three reasons, we suggest that the majority of the observed gas (dominating the velocity) cannot be due to interaction layers, and a hot but quiescent component should exist.

\subsection{Implications for the Hot Halo}
\label{implication}
Based on the discussion in the former three sections, we suggest that the newly discovered high ionization gas likely originates from a volume-filled hot gaseous halo. Compared to other high ionization intervening gas (traced by {\NeVIII}), this {\MgX} system has a similar total column density of $\lesssim 10^{20}\rm ~ cm^2$, but a larger temperature than the previous studies ($\gtrsim 10^6~\rm K$; \citealt{Narayanan:2012aa}).
% In this section, the implication of this hot gas will be discussed.

The mass of a hot gaseous halo is related to the mass of the galaxy, which also defines the virial temperature of the system. Theoretically, it is believed that the massive galaxy can host a hot halo at their virial temperature because of their large gravitational potential and the long cooling timescale, while the virialized gaseous halo of a low mass galaxy will not be formed due to the rapid radiative cooling \citep{Keres:2009aa}. The virialized hot gaseous halo ($\gtrsim 10^6\rm~K$) results in the hot accretion mode, which means the hot gas in the inner $50\rm ~kpc$ radiatively cools and falls onto the galaxy. For temperatures below $5 \times 10^5\rm~ K$, nearly the entire shock-heated halo cools in less than a Hubble time, so a significant hot halo does not form, with the accreting gas being cold \citep{Keres:2009aa}. The only hot halo candidate ($T > 10^6~\rm K$) is in \object{HE 0153-4520} reported by \citet{Savage:2011aa}, based on the presence of a Ly$\alpha$ absorption line with a Doppler $b$ value of $140~\rm km~s^{-1}$. However, without high ionization metal lines to confirm this suggestion, since it is common that the intervening systems show multiple components. The case becomes ambiguous when several weak features (only evident in the wings of the lines) blend with a strong line, because one cannot determine uniquely whether these broadened wings are from a single component.

As discussed in Section \ref{CIM}, a multi-temperature medium is required to model all ions in the {\MgX} system, and the presence of {\MgX} in our sample raises the upper limit of the temperature up to $10^6\rm~K$), compared with the cases with only {\OVI} and {\NeVIII} ($\sim 10^{5.7}\rm~K$). Adopting the power law model, the total hydrogen column density is $8.2\times 10^{19}\rm~cm^{-2}$ at $0.3$ solar metallicity, and the index is $1.55\pm 0.41$, which indicates the higher temperature gas contains most of the mass. Quantitatively, the gas associated with {\MgX} and {\NeVIII} ($5.7 < {\rm log} T \lesssim 6.0$) has a column density of $7.1\times 10^{19}\rm~cm^{-2}$, while the gas corresponded to {\NeVI} and {\OVI} ($5.5 < {\rm log} T < 5.7$) is $7.6\times 10^{18}\rm~cm^{-2}$. This multi-phase medium is similar to the hot halo of the Milky Way, where the {\OVI} gas is one order of magnitude lower than the {\OVII} and {\OVIII} (similar temperature region to {\NeVIII}) gas \citep{Hodges-Kluck:2016aa}. Specifically, the currently observed {\OVII} has EW around $15\rm~m\AA$, which corresponds to a column density of $5.2\times10^{15} \rm~cm^{-2}$ \citep{Nicastro:2002aa, Rasmussen:2003aa}. Assuming an average ionization fraction of $50\%$ and $0.3$ solar metallicity, the total column density is of $5.3\times10^{19}\rm~cm^{-2}$.
%, which is consistent with our case.

Based on our fitting results, the variation of the temperature is $1.65 \rm~dex$, and the characteristic temperature is approximately $10^{5.93}\rm~ K$, which means a half of the hydrogen column density is higher than this temperature. While the hydrogen column density distribution appears to have a peak around {\NeVIII} and {\MgX}, there is no information for temperatures above $\sim 10^{6.3}\rm~ K$. This shortcoming could be removed when sufficiently sensitive X-ray spectrographs can detect ions such as {\OVIII}.

\begin{figure}
\begin{center}
\includegraphics[width=0.48\textwidth]{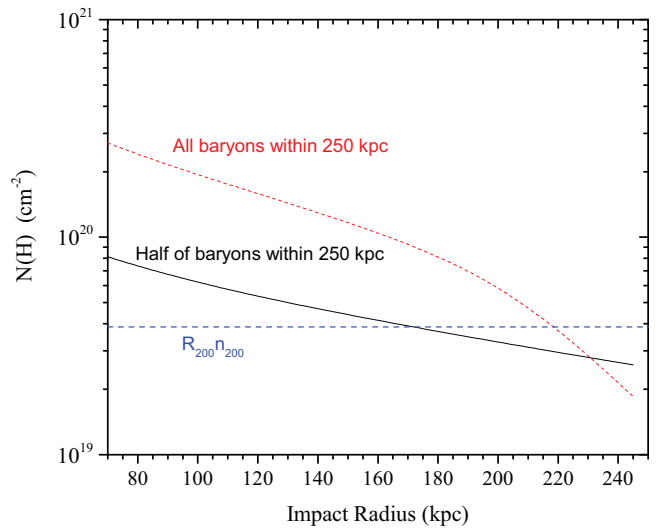}
\end{center}
\caption{For A Milky Way type galaxy, the dotted red line is a model where all the baryons lie within $r = 250\rm~ kpc$, while the black line has half of the baryons within $250\rm~ kpc$ ($R_{200}$). The dashed blue line shows the typical column density $R_{200}n_{200}$.}
\label{f6}
\end{figure}

The total column density of this absorption system is $\sim 8.2 \times 10^{19} (0.3Z_{\odot}/Z)\rm~cm^{-2}$, which could be used to constrain its host. Here, we show estimations of the total column density for three possible hosts (galaxy, galaxy group and galaxy cluster). We assume that the typical halo radii ($R_{200}$) of these three objects are $250\rm~kpc$, $1\rm~Mpc$ and $3\rm~Mpc$, and the mean density is $n_{200}$ ($5\times 10^{-5}\rm~cm^{-3}$). Based on these assumptions, the expected total column densities ($n_{200} R_{200}$) are about $3.9\times 10^{19}$, $1.5\times 10^{20}$ and $4.6\times 10^{20}\rm~cm^{-2}$, respectively. Thus, the column density of the hot gas seems to be reasonable for a galaxy hot halo. For Milky Way like galaxies, we show a detailed model in Fig. \ref{f6}. In this model, we employ a $\beta$-model, where density goes as $r^{-3/2}$ and assume the baryon mass in $250\rm~kpc$ ($50\%$ or $100\%$ of the total baryons). In this plot, it is shown that $n_{200} R_{200}$ is a good indicator for the total column density, and the radial dependence of the column density shows that our absorption system correspond to a projected distance $\sim 80 - 170\rm~kpc$.

With the power law index of $1.55$, it is of interest to consider whether the temperature dependence is due to the local mixture of multiple phase gases or the large scale radial structure of the galaxy halo. Thus, we try to reproduce the power law model in the frame of the galaxy halo. For the radial distribution of the density, we use a modified $\beta-$model described in \citet{Miller:2015aa}. They show that $\beta$ is $0.50\pm0.03$ for the Milky Way, which indicates $n\propto r^{-3/2}$. For the radial distribution of the temperature, we use the result from \citet{Baldi:2012aa} for the outer region of galaxy clusters, where the temperature varies as $T\propto r^{-\alpha}$ with $\alpha$ of $0.6$ to $1.5$. If we assume this law is the same in galaxies or galaxy groups, the relationship between the column density and the temperature is
\begin{equation}
\frac{{\rm d}N}{{\rm d}T} \propto T^{\frac{3\beta-1}{\alpha}-1},
\end{equation}
so the power index is $\lesssim 0$, while our fitting shows a power law of
\begin{equation}
\frac{{\rm d}N}{{\rm d}T} = 10^{4.4\pm2.2 - [Z/X]} T^{1.55\pm0.41},
\end{equation}
where $T$ is in the range $10^{4.39\pm0.13}$ to $10^{6.04\pm0.05} \rm~K$. The difference between the power law index indicates the empirical model is unlikely to be as steep as our observation. This inconsistency shows that the global radial variation does not reproduce the observed ion species. 

The temperature dependence of the column density may be accounted for by the local mixture of multiple phase gases, which have several origins, such as an accretion flow or stellar feedback. The accretion flow can generate cool clouds in the halo, while the stellar feedback can not inject cool clouds to large radii ($\gtrsim 100~\rm kpc$). The involvement of low temperature gas will lead to the mixture with the hot halo, then, the intermediate state can be generated in interaction layers. Therefore, the column density distribution may be divided into three parts -- the cool gas, the interaction layer and the hot halo.

We thank Edmund Hodges-Kluck, Matthew Miller and Jiangtao Li for their helpful discussions on this work. We also thank Blair D. Savage, Bart P. Wakker and the anonymous referee for their thoughtful comments. We are grateful to the members of the COS instrument team for providing the IDL routines COS Tools, and to Jeffrey Scargle for developing Bayesian Blocks and making the code publicly available. The spectral data employed in this paper were obtained from the Mikulski Archive for Space Telescopes (MAST). We greatfully acknowledge financial support from NASA grants NNX15AM93G and NNX16AF23G in support of this work.

\bibliographystyle{apj}
\bibliography{ms}

\end{document}